\documentclass[journal=jacsat,manuscript=letter]{achemso}

\usepackage[version=2]{mhchem} 
\usepackage{graphicx}
\usepackage{color}
\usepackage{verbatim}
\usepackage{setspace}

\author{Nathan Zhao}
\affiliation[Stanford University]
{Department of Applied Physics, Stanford University, Stanford, California 94305, USA}

\author{Zhexin Zhao}
\author{Ian A. D. Williamson}
\affiliation[Second University]
{Department of Electrical Engineering, Stanford University, Stanford, California 94305, USA}

\author{Salim Boutami}
\affiliation[Grenoble]
{Universit\'{e} Grenoble Alpes, CEA, LETI, Grenoble, 38054, France}
\alsoaffiliation[Second University]
{Department of Electrical Engineering, Stanford University, Stanford, California 94305, USA}

\author{Bo Zhao}

\author{Shanhui Fan}
\affiliation[Second University]
{Department of Electrical Engineering, Stanford University, Stanford, California 94305, USA}
\email{shanhui@stanford.edu}

\date{today}

\title[An \textsf{achemso} demo]
  {High reflection from a one-dimensional array of graphene nanoribbons}

\begin{document}

\begin{abstract}
\begin{singlespace}

We show that up to $90\%$ reflectivity can be achieved by using guided plasmonic resonances in a one-dimensional periodic array of plasmonic nanoribbon. In general, to achieve strong reflection from a guided resonance system requires one to operate in the strongly over-coupled regime where the radiative decay rate dominates over the intrinsic loss rate of the resonances. Using an argument similar to what has been previous used to derive the Chu-Harrington limit for antennas, we show theoretically that there is no intrinsic limit for the radiative decay rate even when the system has an atomic scale thickness, in contrast to the existence of such limits on antennas. We also show that the current distribution due to plasmonic resonance can be designed to achieve very high external radiative rate. Our results show that high reflectivity can be achieved in an atomically-thin graphene layer, pointing to a new opportunity for creating atomically-thin optical devices. 
\end{singlespace}

\end{abstract}

\maketitle
\begin{singlespace}

Recently there has been significant interest in achieving strong reflection from atomically thin materials, with potential applications in high efficiency optical modulators \cite{Back2018} and for achieving large optomechanical interactions \cite{Williamson2016}. For this purpose, it is essential to create and utilize various kinds of optical resonances in these materials. For example, it has been recently demonstrated that at low temperature, monolayer MoSe$_2$ can achieve high reflection of incident light due to its excitonic resonance \cite{Back2018,Scuri2018}. 

To achieve strong reflection using a resonance, one must operate in an effectively one-dimensional system where the transmitted and reflected light are restricted to a single diffraction order. Moreover, the resonance must be in the over-coupled regime where the external radiative rate of the resonance dominates over the intrinsic loss rate. Thus it is important to develop a fundamental understanding of the external radiative rate for a resonance in an effective one-dimensional system. For a resonance in a two- or three-dimensional system, such as the resonance found in an antenna, the Chu-Harrington limit constrains the radiative decay rate with an upper bound proportional to the antenna's physical size \cite{Wheeler1947, Chu1948, Harrington1959}.  However, there has not been a similar understanding of whether there exists a fundamental bound on external radiative decay rate for resonances in effective one-dimensional systems. 

In this Letter we theoretically show that there is no upper bound on the radiative decay rate in a one-dimensional resonance. We then demonstrate a practical design approach towards enhancing the radiative decay by engineering the conduction current distribution in a plasmonic resonator consisting of a single-atomic-layer graphene nanoribbons. The resulting structure exhibits high reflection even when realistic loss rates of graphene is taken into account. 

\begin{figure}[b]
  \centering
  \includegraphics[width=4in]{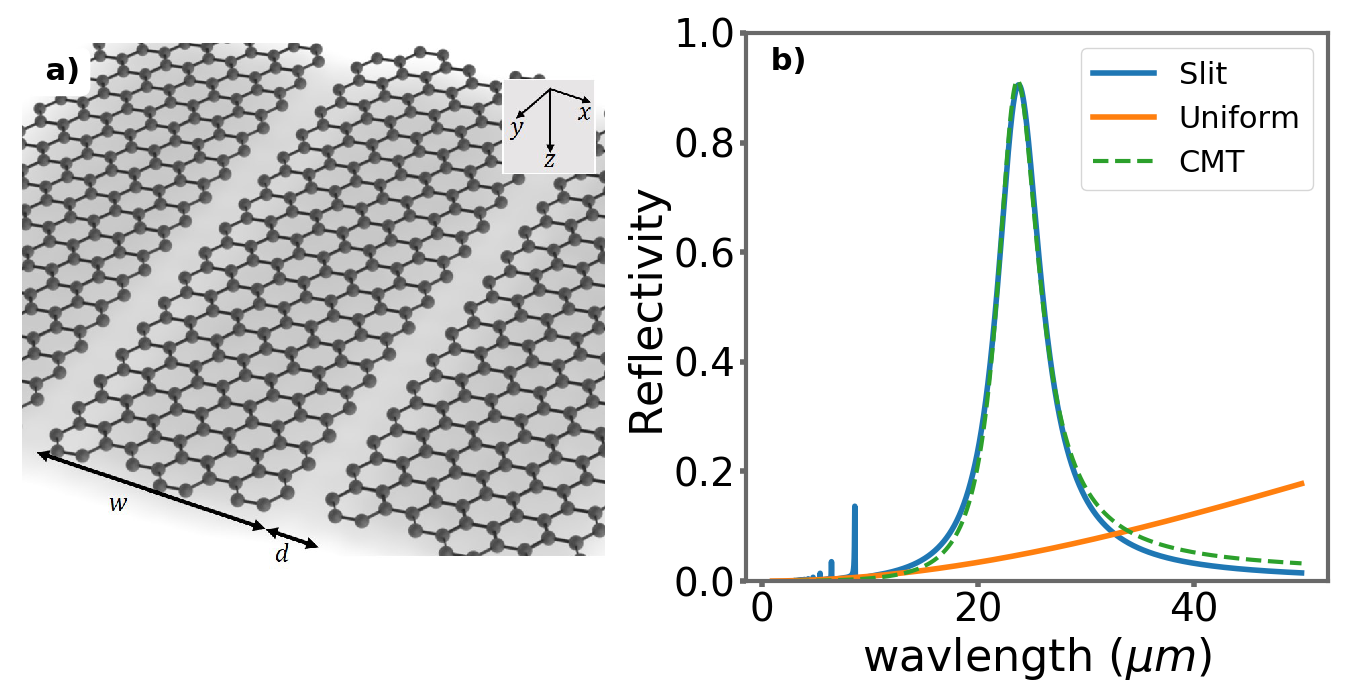}
  \caption{(a) A periodic array of graphene nano-ribbons. b) Numerically computed reflection spectrum for an array of graphene nanoribbons with $w=0.9$ $\mu$m, and $d=0.050$ $\mu$m. The orange line indicates the reflectivity for a single uniform graphene layer and the dashed green line indicates the coupled mode theory (CMT) fit to the $0^{\text{th}}$ order resonance.}
  \label{fig:spectra_sample}
\end{figure}

To understand the role of a resonance in reflection and the need for a large external radiative rate, consider the exemplary geometry as shown in Fig. \ref{fig:spectra_sample}a, where a sheet of graphene nano-ribbons is suspended in air with its reflectivity spectrum shown in Fig. \ref{fig:spectra_sample}b which exhibits strong reflection. We choose the periodicity to be at sub-wavelength scale such that for normally incident light the system behaves effectively as a one-dimensional system. Suppose the system supports a resonance. Then from the temporal coupled mode theory formalism, the reflection of the system has the form
\cite{Fan2003,Haus1984}:
\begin{equation}
    r{\left(\omega\right)} =  e^{j\phi}\bigg(r_b{\left(\omega\right)} - \gamma_r \frac{r_b(\omega) +j t_b(\omega)}{j(\omega-\omega_0 )+\gamma_r+\gamma_i}\bigg),
\label{eq:CMT_r}
\end{equation}
where $\phi$ is a phase factor, $\omega_0$ is the resonant frequency, $r_b$ and $t_b$ are the reflection and transmission of the direct scattering process. $\gamma_r$ represents the external radiative decay rate and $\gamma_i$ represents the internal loss rate. For extremely thin materials such as graphene operating in the mid- to near-infrared, generally $r_b \approx 0$ and $t_b \approx 1$. From Eq. \ref{eq:CMT_r}, high reflectivity requires that the resonance be designed to operate in the over-coupled regime where $\gamma_r \gg \gamma_i$. Therefore, to achieve high reflection it is important to seek to enhance the radiative rate, or equivalently to reduce the quality factor associated with the radiative decay process. 

To design a resonance-based reflector with resonant frequency $\omega_0$, it is therefore important to understand any possible constraint on the radiative decay rate. In two and three-dimensional systems, the radiative decay rate of a resonator is subject to the Chu-Harrington limit. Here we briefly review the arguments of this limit since this understanding is essential for our present work.  Related to the radiative decay rate $\gamma_r$, one can alternatively define a radiative quality factor, $Q_r = \omega_0/(2\gamma_r)$,  which depends on the period-averaged energy stored in the resonator $W$, as well as the period-averaged radiated power $P_{\text{rad}}$ as:
\begin{equation}
\label{eq:Q}    
Q_r = \frac{\omega_0W}{P_{\text{rad}}}.
\end{equation}
Consider a linearly polarized dipole antenna which radiates to free space, and can be bounded by a sphere with radius $r=a$. Assuming that this antenna supports only the TM$_{01}$ mode, then in free space outside the bounding sphere, one of its electric field components has the form \cite{McLean1996}:
\begin{equation}
E_{\theta} \sim e^{-jk_0r} \bigg(-\frac{jk_0}{r}-\frac{1}{r^2} +\frac{j}{k_0r^3} \bigg).
\end{equation}
The first term in the parentheses above corresponds to the radiative field, from which one can determine the total radiative power $P_{\text{rad}}$ in Eq. 2. The second and third terms correspond to the non-radiative near-field. Integrating the energy for such near-field component in the volume outside the bounding sphere, we get a lower bound on the stored energy. Therefore, from Eq. 2 one obtains a lower bound of the radiative quality factor for the dipole antenna, 
\begin{equation}
Q_r \geq \frac{1}{k_0a}+\frac{1}{(k_0a)^3}.
\end{equation}
While the derivation here is for a dipole antenna, one can in fact show that this bound applies in general for any antenna \cite{McLean1996}. A similar derivation can be carried out for two-dimensional systems \cite{Collin1964}. 

The essence of the derivation above is that in the spherical coordinate system which is appropriate for three dimensions, an outgoing wave in free space always contains near-field components, as shown in Eq. 3. And hence there is always energy storage associated with such an outgoing wave. Such energy storage necessitates a lower bound in the radiative quality factor. On the other hand, for a one-dimensional system, an outgoing wave solution in free space has the form:
\begin{equation}
E_{x}(z) \sim e^{-jk|z|},
\end{equation}
which need not have any near-field component. Thus, using the same argument for the Chu-Harrington limit as discussed above, one should conclude that there is no limit on the lower bound of the radiative quality factor for a one-dimensional system. 

We now show both analytically and numerically that the structure as shown in Fig. 1a, which consists of an array of suspended graphene nanoribbons, provides a pathway to achieve resonances with very high radiative rate. Numerically, we use the Rigorous Coupled Wave Analysis (RCWA) to simulate the structure shown in Fig. 1a. We describe the conductivity of graphene as:
\begin{equation}
\begin{split}
\sigma(\omega)=\frac{2ie^2 k_b T}{\pi\hbar(\omega+i\tau^{-1})}\ln{\big[2\cos(\frac{\mu}{2k_b T}) \big]} + \\
\frac{e^2}{4\hbar} \bigg( G{(\omega/2)}- \frac{4\omega}{i\pi} \int_{0}^{\infty} d\epsilon \frac{G(\epsilon)-G(\omega/2)}{\omega^2-4\epsilon^2} \bigg),
\end{split}
\label{eq:graphene_cond}
\end{equation}
where the first term is the intraband term and the second is the interband term and $G{(\epsilon)} = \sinh{(\epsilon/(k_bT))}/(\cosh{(\mu/(k_bT))}+\sinh{(\epsilon/(k_bT))})$, with $k_b$ the Boltzmann constant and $T$ the temperature \cite{Falkovsky2008, Hanson2008}.  In Eq. \ref{eq:graphene_cond}, $\epsilon$ is the electron energy (in the conduction band), $\mu$ is the chemical potential, $\tau$ is the scattering time. Unless otherwise noted, in this paper, we choose $\mu = 0.8$ eV, $\tau = 1.25 \times 10^{-12}$ s (or an approximate mobility of $10000$ $\frac{\text{cm}^2}{\text{V} \cdot \text{s}}$ and carrier density of $ 5 \times 10^{13}$ cm$^{-2}$) to minimize the intrinsic loss while staying close to known experimental results \cite{Gao2012,Craciun2011,Bolotin2008, Dean2010,Banszerus2015, Yin2015, Ye2011, Efetov2010, Ni2018}. In the RCWA simulation, the graphene sheet is modeled as an effective dielectric layer with a thickness ($h$) of $0.34$ 
nm with a frequency dependent dielectric constant \cite{Vakil2011}:
\begin{equation}
  \epsilon_r(\omega)=1+j\bigg(\frac{\sigma(\omega)}{\omega\epsilon_0 h} \bigg).
\label{eq:graphene_eps}
\end{equation}

A uniform graphene sheet supports plasmons which are TM-polarized, with the non-zero field components being $H_y$, $E_x$, and $E_z$. In the structure of Fig. 1a, the periodicity along the $x$-direction causes some of these plasmons to radiate into the free space, creating a guided resonance. Here we consider only normally incident light with $k_x = 0$, and choose the periodicity to be below the free space wavelength of light such that the system behaves as an effective one-dimensional system. 

For an analytic treatment of the radiative rate $\gamma_r$, we must relate the radiated power to specific features of the graphene plasmonic resonator. Such a resonator is described by the surface current density $J_x(x)$. From the surface boundary condition
\begin{equation}
    \label{eq:sbc}
   J_x{\left(x\right)} = -\left[ H_y{\left(x,z= 0^+\right)} - H_y{\left(x,z= 0^-\right)} \right].
\end{equation}
and Maxwell's equations, we can use $J_x(x)$ to determine all the fields of the resonance. Moreover, the structure in Fig. 1a is periodic with mirror symmetry about $x=0$. Therefore, we can decompose the surface current as a Fourier series:
\begin{equation}
\label{eq:fourier_decomp}
J_x(x) = J_x^{(0)} + \sum\limits_{n=1}^{\infty}J_x^{(n)} \cos{\left(\frac{2\pi n}{w} x\right)}.
\end{equation}
The radiated power $P_{\text{rad}}$ is only dependent on the 0$^{\text{th}}$ order Fourier component:
\begin{equation}
 P_{\text{rad}}=\int_{-a/2}^{+a/2} \text{Re}{\left( E_x^{(0)} {H_y^{(0)}}^* \right)} dx = \frac{w}{4} \sqrt{\frac{\mu_0}{\epsilon_0}} \left|J_x^{(0)}\right|^2.
 \label{eq:power}
\end{equation}
Here we take into account that the radiation can go both upward and downward. The higher order components of the Fourier decomposition in Eq. \ref{eq:fourier_decomp} contribute to the stored energy $W$, which include both the energy stored in the electromagnetic field, as well as in the kinetic energy of the electrons as described in terms of a kinetic inductance \cite{Staffaroni2012}. From Eq. \ref{eq:Q}, minimizing the $Q_r$ can be framed as maximizing the ratio of $P_{\text{rad}}$ to $W$. With Eq. \ref{eq:power}, we can now see that to minimize the $Q_r$, one must maximize the relative contribution of $J_x^{(0)}$ compared to the higher order components $J_x^{(n)}$, $n>0$. We thus define the relative contribution of the $k^{th}$ component as:
\begin{equation}
    R_{k} = \frac{\big|J_x^{(k)}\big|^2(1+\delta_{0k})}{\sum_{n=0}^{N}\big(\big|J_x^{(n)}\big|^2(1+\delta_{0n})\big)},
\label{eq:relative_contr}
\end{equation}
where $\delta_{0k}$ is the kronecker delta. Eq. \ref{eq:relative_contr} accounts for the space averaging of the cosine in all the higher order components. The numerically determined surface current distribution $J_x(x)$, for a few nanoribbon array structures, is shown in Fig. 2 (The RCWA simulations provide the magnetic field distributions. The surface current distributions are then obtained using Eq. (8)). A prominent feature of the current distribution is the presence of a kink, i.e. a discontinuity in its first derivative, at the edge of the ribbons. This kink, moreover, persists even when the air gap between the ribbon shrinks in size.  Such a kink is related to the diverging charge density at the edges of the ribbon. From the charge conservation equation, $\nabla \cdot \textbf{J}= -i\omega \rho$, within the graphene sheet at its edge, we have $dJ_x/dx \to \infty$, and hence $dx/dJ_x \to 0$.  We can then perform a Taylor expansion of $x(J_x)$ around $x=\pm w/2$ as:
\begin{equation}
 x(J_x ) \approx \frac{w}{2} \pm \frac{1}{2}  \bigg(\frac{d^2 x}{d J_x^2}\bigg) J_x^2 + \ldots
 \label{eq:Taylor_J}
\end{equation}
\begin{figure*}
  \centering
  \includegraphics[width=110mm]{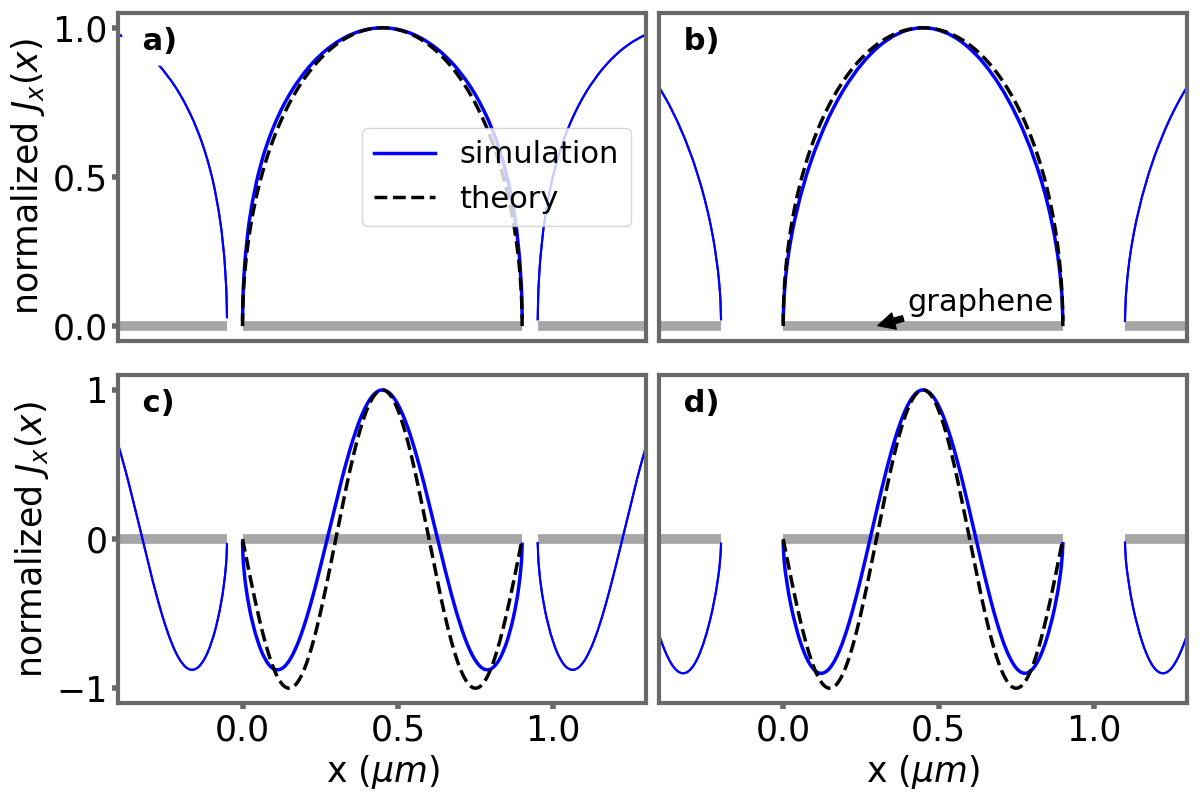}  
  \caption{Normalized distributions of the $x$-component of the current for an array of graphene ribbons with width $w=0.9$ $\mu$m and various gap sizes $d$. The gray bars represent the graphene nanoribbons. (a) $d = 50$ nm, the lowest order mode. (b) $d =$ 200 nm, the lowest order mode. (c) $d = $ 50 nm, the 2nd order mode. (d) $d = $  200 nm, the second order mode. }
  \label{fig:current_view}
\end{figure*}
Thus, near the edges at $x = \pm w/2$ we have $J_x(x) \sim
(x\pm w/2)^{1/2}$. In Ref. \citenum{Peres2013}, by interpolating the behavior of $J_x(x)$ between two edges, it was argued that $J_x(x)$ has the following form for the lowest order resonance:
\begin{equation}
\label{eq:current}
  J_x{\left(x\right)} =
  \begin{cases}
        \frac{2}{w}\sqrt{(\frac{w}{2})^2-x^2} & \text{if } |x|\leq\frac{w}{2} \\
                                   0 & \text{if } \frac{w}{2}<|x|<a  .
  \end{cases}
\end{equation}
For our system. Eq. \ref{eq:current} agrees quite well to the numerically determined current distribution over a wide range of gap sizes shown in Fig. \ref{fig:current_view}. 

\begin{figure} [!hbtp]
  \centering
  \vspace{-0.5cm}
  \includegraphics[width=120mm]{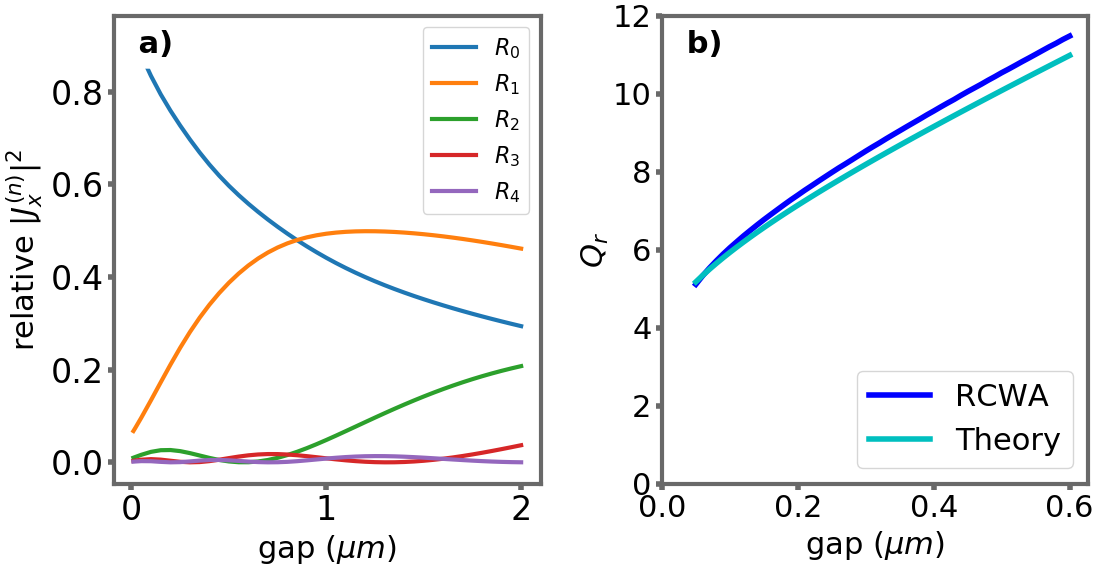}
  \caption{(a) Relative contribution ($R_n$ in Eq. \ref{eq:relative_contr}) of the Fourier components of the surface current function $J_x(x)$ in Eq. \ref{eq:current} as a function of gap size. The systems consist of an array of ribbons with $w=0.90$ $\mu$m. (b) The cyan curve is the calculated $Q_r$ using the fields derived from the analytic ansatz in Eq. \ref{eq:current}, the blue line is the $Q_r$ determined from our RCWA reflection spectra as fitted using coupled mode theory. The ribbon has a width of $w = 0.90$ $\mu$m.}
  \label{fig:fitted_Q}
\end{figure}

Given $J_x(x)$ in Eq. \ref{eq:current}, we can now decompose it in a Fourier series and analyze how the geometric parameters, $d$ and $w$, affect the relative contribution of $J_x^{(0)}$. From Fig. \ref{fig:fitted_Q}a, the relative contribution of the $0^{\text{th}}$ component of $J_x(x)$ increases rapidly with decreasing gap size $d$ for fixed $w$, while the relative contributions of all the higher order components decrease with decreasing gap size. This clearly indicates that the ratio of the power radiated relative to the power stored is increasing. Consequently, the predicted $Q_r$, calculated with Eq. \ref{eq:Q}, \ref{eq:sbc}, \ref{eq:fourier_decomp}, and \ref{eq:power} decreases with decreasing gap size according to our theory, as shown by the cyan line in Fig. \ref{fig:fitted_Q}b.

To support our analysis, we numerically compute the $Q_r$ using the rigorous coupled wave analysis of the same structure. For the same set of structures analyzed in Fig. \ref{fig:fitted_Q}b, we simulate their reflection spectra. An exemplary spectrum, for the structure with the width of ribbon $w = 0.9$ $\mu$m, and gap size $d = 0.050$ $\mu$m, is shown in Fig. \ref{fig:spectra_sample}b. The spectrum features several peaks, and we focus first on the lowest order resonance which has the longest wavelength, for which the theory as developed above is applicable. To determine its radiative quality factor $Q_r$, we fit the reflectivity spectrum using Eq.\ref{eq:CMT_r}1. As shown in Fig. \ref{fig:spectra_sample}b, the fit agrees quite well with the numerically determined reflection spectrum. The radiative quality factor $Q_r$, thus determined for varying gap sizes, is plotted in Fig. \ref{fig:fitted_Q}b as the blue line, which agrees quite well with the analytic prediction. Thus, we have indeed shown that very low radiative quality factor, down to the single digits, i.e. a very high radiative rate, is achievable in this structure as the gap size reduces. The fit to the reflectivity spectra also determines the intrinsic loss rate $Q_i$. For the set of structures considered in Fig. \ref{fig:fitted_Q}b, $Q_i \approx 100$ is more than an order of magnitude higher than $Q_r$. Thus the structures are in the over-coupled regime and exhibits strong reflectivity at resonance. The structure with a gap size of $0.050$ $\mu$m has a peak reflectivity of $90.5\%$ as shown in Fig. \ref{fig:spectra_sample}b. Thus, we have shown that high reflectivity can be achieved in the graphene nanoribbon array which is atomically thin. 

The spectrum in Fig. \ref{fig:spectra_sample}b also exhibits narrower peaks at shorter wavelengths which correspond to higher-order resonances. Fig. 2c and d shows the current distribution for the second-order resonance. This resonance is the next higher-order resonance that has an even mirror symmetry with respect to the center of the graphene ribbon. Having such an even symmetry is necessary in order for the mode to couple to external radiation from normal incidence. However, the current distribution of this resonance closely resembles a sinusoidal function. Such oscillation of the current distribution for this mode inevitably leads to much larger higher-order Fourier components in Eq. (9) and hence a much lower radiative rate.

\begin{figure}[!htbp]
  \centering
  \includegraphics[width=75mm]{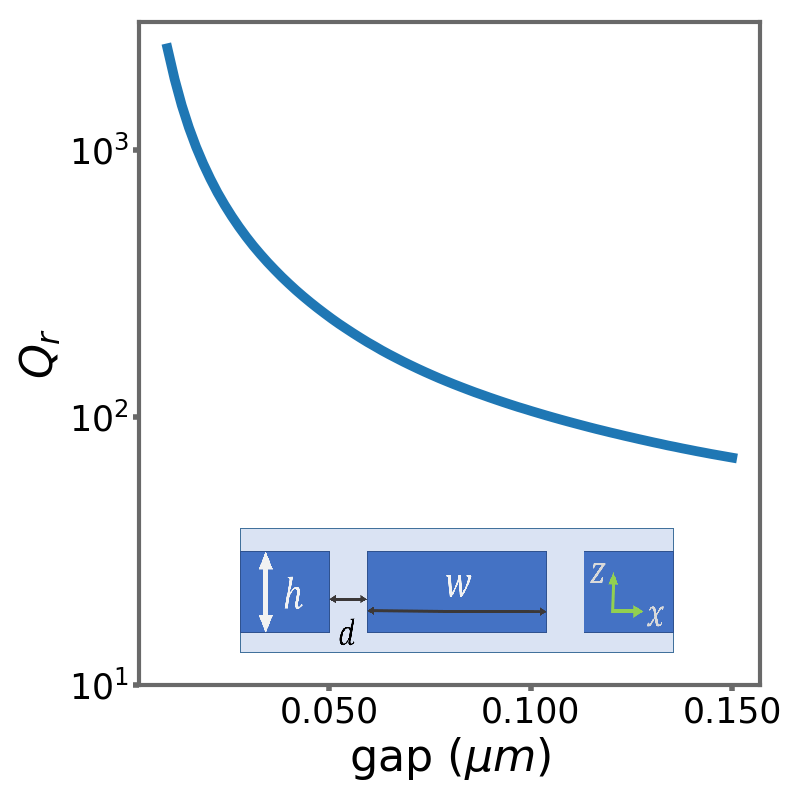}
  \caption{Numerically computed radiative Q-factor for the dielectric grating system. For the dielectric grating (shown in the inset), we use a periodicity along the $x$-direction of 0.76 $\mu$m. The grating has a thickness $h=0.25$ $\mu$m and a dielectric constant of $3.5$. Here, we consider the lowest order guided resonance with a wavelength of approximately 0.96 $\mu$m.}
  \label{fig:dielectric_compare}
\end{figure}

The behavior where the radiative linewidth increases as the gap size decreases was previously observed experimentally in metallic grating structures\cite{Smythe2007} but not theoretically explained. While the present focus of the paper is on graphene resonators, our theory also provides a theoretical explanation of the experimental results in Ref. 25. Such behavior is unique to plasmonic systems like the nano-ribbon array and does not occur in an all-dielectric guided resonance system. As an illustration, in Fig. \ref{fig:dielectric_compare} we consider a dielectric grating structure with a periodic array of air slits introduced into a dielectric slab waveguide. Such a system supports guided resonances\cite{Fan2002}. The radiation rate of the guided resonance  decreases as the gap sizes decreases since the lateral ($x$-direction) profile of the guided resonance smoothly approaches that of the guided mode of the dielectric waveguide as the gap size decreases.

\begin{figure}
  \centering
  \vspace*{-0.7 cm}
  \hspace{0.2 cm}
  \includegraphics[width=100mm]{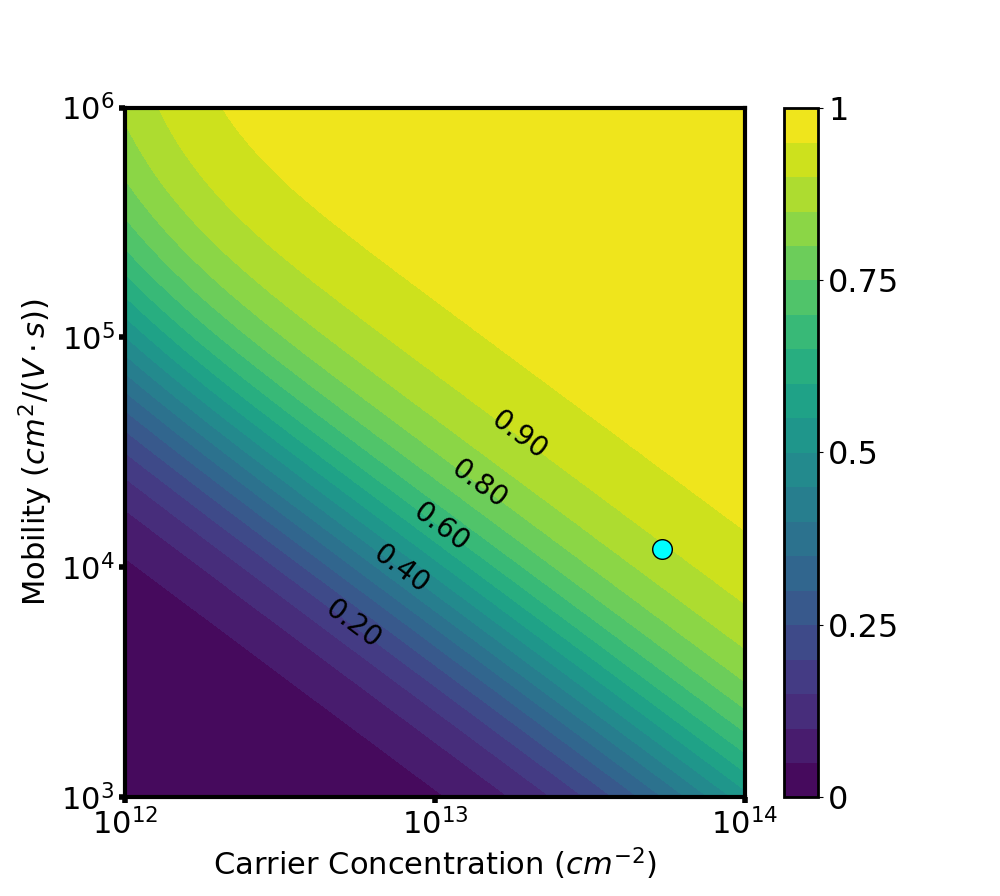}  
  \caption{Color map of the peak resonant reflectivity of an array of graphene nanoribbons with $w=0.9$ $\mu$m and $d=50$ nm and varying combinations of mobility and carrier concentration. The blue dot indicates the set of parameters used for Fig. 1b.}
  \label{fig:graphene_params}
\end{figure}

Before concluding, we briefly discuss the factors that control the internal loss rate $\gamma_i$. Unlike the external radiative decay rate $\gamma_r$, which is strongly structure dependent, the internal loss rate $\gamma_i$ is only weakly dependent on the structural geometry, and is instead mostly controlled by material parameters, such as the chemical potential $\mu$ and the scattering time $\tau$ in Eq. \ref{eq:graphene_cond}. These parameters are related to the carrier concentration and the mobility, both of which are more accessible experimentally. The carrier concentration can be tuned via electrostatic gating and/or doping \cite{Craciun2011, Pachoud2010}, whereas the mobility can be directly measured. In Fig. \ref{fig:graphene_params}, we show the dependency of the peak reflectivity on carrier concentrations and mobility, for the structure shown in Fig. \ref{fig:spectra_sample}b. We see a strong dependency of the reflectivity on these parameters. To achieve high reflectivity generally requires high carrier concentration and high mobility. The choice of the parameters for the spectrum shown in Fig. 1b, as indicated by a cyan dot in Fig. \ref{fig:graphene_params}, reflects this requirement, as well as the trade-off between optimizing mobility versus increasing carrier concentration. While the focus of the paper is on single layer graphene, we note that higher mobility and carrier concentrations can be achieved in bi-layer or tri-layer graphenes \cite{Ye2011, Zhu2009}, which maybe more favorable for achieving high reflection.

In conclusion, we have shown that a periodic array of graphene nanoribbon can be designed to achieve high reflectivity. The underlying concept relies upon the general observation of the lack of Chu-Harrington limit in one-dimensional systems, and the unique current distribution in graphene nanoribbons. Such high reflectivity, in combination with other aspects of graphene, such as large in-plane Young’s modulus, high melting point,  may open up opportunities for reflectors \cite{Williamson2016, Sun2018, Ginzburg2013}, terahertz antennas \cite{Perruisseau-Carrier2012,Zayats2013}, and potentially light sails \cite{Atwater2018}. 

This work is supported by an U. S. AFOSR MURI project (Grant No.  FA9550-17-1-0002). 

\providecommand{\latin}[1]{#1}
\makeatletter
\providecommand{\doi}
  {\begingroup\let\do\@makeother\dospecials
  \catcode`\{=1 \catcode`\}=2\doi@aux}
\providecommand{\doi@aux}[1]{\endgroup\texttt{#1}}
\makeatother
\providecommand*\mcitethebibliography{\thebibliography}
\csname @ifundefined\endcsname{endmcitethebibliography}
  {\let\endmcitethebibliography\endthebibliography}{}

\end{singlespace}

\end{document}